\documentclass[article]{IEEEtran}

\usepackage[english]{babel}
\usepackage[latin1]{inputenc}
\usepackage{enumerate}
\usepackage{color}
\usepackage[T1]{fontenc}
\usepackage{subfigure}
\usepackage{dsfont}
\usepackage{graphicx}
\usepackage{epstopdf}
\DeclareGraphicsExtensions{.eps}
\usepackage[T1]{fontenc}
\usepackage{amsmath}
\usepackage{mathtools}
\usepackage{amsthm}
\usepackage{amstext}
\usepackage{amssymb}
\usepackage{mathrsfs}
\usepackage{cite}
\usepackage{mathtools}
\usepackage{tikz}
\usetikzlibrary{arrows,positioning}

\def\R{\mathbb{R}}

\def\Pr{\mathop{\rm Pr}}

\def\B{{\mathcal B}}

\def\P{{\mathcal P}}

\def\S{{\mathcal S}}

\def\cE{\mathbb{E}}

\def\sPr{{\mathsf{Pr}}}

\def\sX{{\mathsf X}}
\def\sY{{\mathsf Y}}
\def\sA{{\mathsf A}}
\def\sH{{\mathsf H}}
\def\sZ{{\mathsf Z}}

\def\sE{{\mathsf E}}

\def\sV{{\mathsf V}}
\def\sW{{\mathsf W}}

\theoremstyle{remark}
\newtheorem{example}{Example}

\newtheorem{theorem}{Theorem}

\theoremstyle{remark}
\newtheorem{remark}{Remark}
\newtheorem{assumption}{Assumption}

\IEEEoverridecommandlockouts

\allowdisplaybreaks

\begin{document}
\sloppy
\title{Finite Model Approximations for Partially Observed Markov Decision Processes with Discounted Cost
\thanks{This research was supported in part by the Natural Sciences and Engineering Research Council (NSERC) of Canada.}
}
\author{Naci Saldi, Serdar Y\"uksel, Tam\'{a}s Linder
\thanks{N. Saldi is with the Department of Natural and Mathematical Sciences, Ozyegin University, Cekmekoy, Istanbul, Turkey, S. Y\"uksel and T. Linder are with the Department of Mathematics and Statistics, Queen's University, Kingston, ON, Canada. Email: naci.saldi@ozyegin.edu.tr, yuksel@mast.queensu.ca, linder@mast.queensu.ca.}
     }
\maketitle

\begin{abstract}
We consider finite model approximations of discrete-time partially observed Markov decision processes (POMDPs) under the discounted cost criterion. After converting the original partially observed stochastic control problem to a fully observed one on the belief space, the finite models are obtained through the uniform quantization of the state and action spaces of the belief space Markov decision process (MDP). Under mild assumptions on the components of the original model, it is established that the policies obtained from these finite models are nearly optimal for the belief space MDP, and so, for the original partially observed problem. The assumptions essentially require that the belief space MDP satisfies a mild weak continuity condition. We provide examples and introduce explicit approximation procedures for the quantization of the set of probability measures on the state space of POMDP (i.e., belief space).
\end{abstract}

\section{Introduction}\label{sec0}

In POMDP theory, existence of optimal policies have in general been established via converting the original partially observed stochastic control problem to a fully observed one on the belief space, leading to a belief-MDP. However, computing an optimal policy for this fully observed model, and so for the original POMDP, using well known dynamic programming algorithms is challenging even if the original system has finite state and action spaces, since the state space of the fully observed model is always uncountable. One way to overcome this difficulty is to compute an approximately optimal policy instead of a true optimal policy by constructing a reduced model for the fully observed system for which one can apply well known algorithms such as policy iteration, value iteration, and $Q$-learning etc. to obtain the optimal policy.

In MDP theory, various methods have been developed to compute near optimal policies by reducing the original problem into a simpler one. A partial list of these techniques is as follows: approximate dynamic programming, approximate value or policy iteration, simulation-based techniques, neuro-dynamic programming (or reinforcement learning), state aggregation, etc. We refer the reader to \cite{Fox71,Whi78,Whi79,Lan81,BeTs96,ReKr02,Ort07,Whi80,Whi82,Ber75,DuPr13,DuPr14} and references therein. However, existing works mostly study systems with discrete (i.e., finite or countable) state and action spaces \cite{Fox71,Roy06,Whi80,Whi82,Cav86,Ort07}) or those that consider general state and action spaces (see, e.g., \cite{DuPr12,DuPr13,DuPr14,Ber75,chow1991optimal}) assume in general Lipschitz type continuity conditions on the transition probability and the one-stage cost function in order to provide a rate of convergence analysis for the approximation error. However, for the fully observed reduction of POMDP, a Lipschitz type regularity condition on the transition probability is in general prohibitive. Indeed, demonstrating even the arguably most relaxed regularity condition on the transition probability (i.e., weak continuity in state-action variable), is a challenging problem as was recently demonstrated in \cite{FeKaZa12} for general state and action spaces (see also \cite{budhiraja2002invariant} for a control-free setup). Therefore, results developed in prior literature cannot in general be applied to compute approximately optimal policies for fully observed reduction of POMDP, and so, for the original POMDP.

In \cite{SaYuLi16,SaYuLi17} we investigated finite action and state approximations of fully observed stochastic control problems with general state and action spaces under the discounted cost and average cost optimality criteria. For the discounted cost case, we showed that optimal policies obtained from these finite models asymptotically achieve the optimal cost for the original problem under the weak continuity assumption on the transition probability. Here, we apply and properly generalize the results in these papers to obtain approximation results for fully observed reduction of POMDPs, and so, for POMDPs. The versatility of approximation results under weak continuity conditions become particularly evident while investigating the applicability of these results to the partially observed case.

In the literature there exist various, mostly numerical and computational, results for obtaining approximately optimal policies for POMDPs. In the following, we list a number of such related results and comparisons with our paper: (i) Reference \cite{porta2006point} develops a computational algorithm, utilizing structural convexity properties of the value function of belief-MDPs, for the solutions of POMDPs when the state space is continuous and action and measurements are discrete, and with further extensions to continuous action and measurements. Reference \cite{spaan2005perseus} provides an algorithm which may be regarded as a quantization of the belief space. However, no rigorous convergence results are given regarding this computational algorithm. (ii) References \cite{smith2012point} and \cite{pineau2006anytime} present quantization based algorithms for the belief state, where the state, measurement, and the action sets are finite. (iii) References \cite{zhou2008density} and \cite{zhou2010solving} provide an explicit quantization method for the set of probability measures containing the belief states, where the state space, unlike in many other contributions in the literature, is continuous. The quantization is done through the approximations as measured by Kullback-Leibler divergence: Kullback-Leibler divergence (or relative entropy) is a very strong pseudo-distance measure which is even stronger than total variation (by Pinsker's inequality \cite{GrayInfo}), which in turn is stronger than weak convergence. In particular, being able to quantize the space of probability measures with finitely many balls as defined by such a distance measure requires very strict assumptions on the allowable beliefs and it in particular requires, typically equicontinuity conditions (see e.g. \cite[Lemma 4.3]{YukselOptimizationofChannels}). (iv) In \cite{YuBe08} the authors consider the near optimality of finite-state controllers that are finite-state probabilistic automatons taking observations as inputs and producing controls as the outputs. A special case for these type of controllers are the ones that only use finite observation history. A similar finite memory approximation is developed in \cite{WhSc94}.  (v) In \cite{YuBe04} the authors establish finite state approximation schemes for the belief-MDPs under both discounted cost and average cost criteria using concavity properties of the corresponding value function and show that approximate costs can be used as lower bounds for the optimal cost function. A similar finite state approximation is considered in \cite{Lov91,ZhHa01} using concavity and convexity properties of the value function for the discounted cost criterion.
We refer the reader to the survey papers \cite{Lov91-(b),Whi91} and the book \cite{Kri16} for further algorithmic and computational procedures for approximating POMDPs.

\textbf{Contributions of the paper.} (i) We show that finite models asymptotically approximate the original partially observed Markov decision process (POMDP) in the sense that the true costs of the policies obtained from these finite models converge to the optimal cost of the original model. The finite models are constructed by discretizing both the state and action spaces of the equivalent fully-observed belief space formulation of the POMDP. We establish the result for models with general state and action spaces under mild conditions on the system components. (ii) We provide systematic procedures for the quantization of the set of probability measures on the state space of POMDPs which is the state space of belief-MDPs. (iii) Our rigorous results can be used to justify the novel quantization techniques presented in \cite{spaan2005perseus,smith2012point,pineau2006anytime} as well as a more relaxed version for the results presented in \cite{zhou2008density} and \cite{zhou2010solving}. In particular, there do not exist approximation results of the generality presented in our paper with asymptotic performance guarantees. We show that provided that the belief space is quantized according to balls generated through metrics that metrize the weak convergence topology (such as Prokhorov, bounded-Lipschitz, or stronger ones such as the Wasserstein metric), and provided that the action sets are quantized in a uniform fashion, under very weak conditions on the controlled Markov chain (namely the weak continuity of the kernel and total variation continuity of the measurement channel), asymptotic optimality is guaranteed. (iv) Our approach also highlights the difficulties of obtaining explicit rates of convergence results for approximation methods for POMDPs. Even in fully observed models, for obtaining explicit rates of convergence, one needs strong continuity conditions of the Lipschitz type, e.g.; \cite[Theorem 5.1]{SaYuLi17}. As Theorem \ref{weak:thm6} shows, this is impossible under even quite strong conditions for POMDPs.



The rest of the paper is organized as follows. In Section~\ref{sec1} we introduce the partially observed stochastic control model and construct the belief space formulation. In Section~\ref{sec2} we establish the continuity properties that are satisfied by the transition probability of the belief space MDP and state approximation results. In Section~\ref{example} we illustrate our results by considering numerical examples. Section~\ref{conc} concludes the paper.

\section{Partially Observed Markov Decision Processes}\label{sec1}

A discrete-time partially observed Markov decision process (POMDP) \index{Partially observed Markov decision process (POMDP)} has the following components: (i) State space $\sX$, action space $\sA$, and observation space $\sY$, all Borel spaces, (ii) $p(\,\cdot\,|x,a)$ is the transition probability of the next state given the current state-action pair is $(x,a)$, (iii) $r(\,\cdot\,|x)$ is the observation channel giving the probability of the current observation given the current state variable $x$, and (iv) the one-stage cost function $c:\sX \times \sA\rightarrow[0,\infty)$.

The key difference of POMDPs from the fully observed MDP is that the state $x_t$ of the system cannot be observed directly by the controller. Instead, the noisy version $y_t$ of the state is available to the controller through an observation channel $r(\,\cdot\,|x_t)$ or $r(\,\cdot\,|x_t, a_t)$. These type of problems in general arise when there is an uncertainty in measurements of the states or there are some states which cannot be measured.

To complete the description of the partially observed control model, we must specify how the controller designs its control law at each time step. To this end, define the history spaces $\sH_0=\sY$ and $\sH_{t}=(\sY\times\sA)^{t}\times\sY$, $t=1,2,\ldots$ endowed with their product Borel $\sigma$-algebras generated by $\B(\sY)$ and $\B(\sA)$. A \emph{policy} $\pi=\{\pi_{t}\}$ is a sequence of stochastic kernels on $\sA$ given $\sH_{t}$. We denote by $\Pi$ the set of all policies.

According to the Ionescu Tulcea theorem \cite{HeLa96}, an initial distribution $\mu$ on $\sX$ and a policy $\pi$ define a unique probability measure $P_{\mu}^{\pi}$ on $\sH_{\infty} \times \sX^{\infty}$. The expectation with respect to $P_{\mu}^{\pi}$ is denoted by $\cE_{\mu}^{\pi}$. For any initial distribution $\mu$ and policy $\pi$ we can think of the POMDP as a stochastic process $\bigl\{ x_t,y_t,a_t \bigr\}_{t\geq0}$ defined on the probability space $\bigl( \Omega, {\cal F}, P_{\mu}^{\pi} \bigr)$, where $\Omega = \sH_{\infty} \times \sX^{\infty}$, the $x_t$ are $\sX$-valued random variables, the $y_t$ are $\sY$-valued random variables, the $a_t$ are $\sA$-valued random variables, and they satisfy for all $t\geq1$
\begin{align}
&P_{\mu}^{\pi}(x_0\in\,\cdot\,)=\mu(\,\cdot\,)\nonumber \\*
&P_{\mu}^{\pi}(x_t\in\,\cdot\,|x_{\{0,t-1\}},y_{\{0,t-1\}},a_{\{0,t-1\}}) \nonumber \\
&\phantom{xxxxxxxxxxx}=P_{\mu}^{\pi}(x_t\in\,\cdot\,|x_{t-1},a_{t-1})=p(\,\cdot\,|x_{t-1},a_{t-1}) \nonumber \\
&P_{\mu}^{\pi}(y_t\in\,\cdot\,|x_{\{0,t\}},y_{\{0,t-1\}},a_{\{0,t-1\}}) \nonumber \\
&\phantom{xxxxxxxxxxx}=P_{\mu}^{\pi}(y_t\in\,\cdot\,|x_{t})=r(\,\cdot\,|x_{t}) \nonumber \\
&P_{\mu}^{\pi}(a_t\in\,\cdot\,|x_{\{0,t\}},y_{\{0,t\}},a_{\{0,t-1\}})=\pi_t(\,\cdot\,|y_{\{0,t\}},a_{\{0,t-1\}}) \nonumber
\end{align}
where $x_{\{0,t\}}=(x_0,\ldots,x_t)$, $y_{\{0,t\}}=(y_0,\ldots,y_t)$, and $a_{\{0,t\}}=(a_0,\ldots,a_t)$. We denote by $J(\pi,\mu)$ the discounted cost function of the policy $\pi \in \Pi$ with initial distribution $\mu$, which is given by \begin{align}
J(\pi,\mu) \coloneqq \cE_{\mu}^{\pi} \biggl[\sum_{t=0}^{\infty} \beta^t  c(x_t,a_t) \biggr], \nonumber
\end{align}
where $\beta \in (0,1)$ is the discount factor.

With this notation, the discounted value function of the control problem is defined as
\begin{align}
J^*(\mu) \coloneqq \inf_{\pi \in \Pi} J(\pi,\mu). \nonumber
\end{align}
A policy $\pi^*$ is said to be optimal if $J(\pi^*,\mu) = J^*(\mu)$.

In POMDPs, since the information available to the decision maker is a noisy version of the state, one cannot apply the dynamic programming principle directly as the one-stage cost function depends on the exact state information. A canonical way to overcome this difficulty is converting the original partially observed control problem to a fully observed one on the belief space. Indeed, let us define
\begin{align}
z_t(\,\cdot\,) \coloneqq \sPr\{x_{t} \in \,\cdot\, | y_0,\ldots,y_t, a_0, \ldots, a_{t-1}\} \in \P(\sX). \nonumber
\end{align}
Here, $z_t$ is the posterior state distributions or 'beliefs` of the observer at time $t$ in the original problem. One can prove that, for any $t\geq0$, we have
\begin{align}
\sPr\{z_{t+1} \in \,\cdot\, | z_0,\ldots,z_t, a_0, \ldots, a_{t}\} &= \sPr\{z_{t+1} \in \,\cdot\, | z_t, a_{t}\} \nonumber \\
&= \eta(\,\cdot\,|z_t,a_t), \nonumber
\end{align}
where $\eta$ is a fixed stochastic kernel on $\P(\sX)$ given $\P(\sX)\times\sA$ (see next section for the construction of $\eta$). Note that $\sPr\{z_{0} \in \,\cdot\,\} = \delta_{\mu}(\,\cdot\,)$.
Furthermore, define the one-stage cost function $\tilde{c}:\P(\sX)\times\sA \rightarrow[0,\infty)$ as
\begin{align}
\tilde{c}(z,a) \coloneqq \int_{\sX} c(x,a) z(dx). \nonumber
\end{align}
Hence, we obtain a fully-observed Markov decision process with the components $(\P(\sX),\sA,\eta,\tilde{c})$. This MDP is called the belief-MDP and it is equivalent to the original POMDP in the sense that for any optimal policy for the belief-MDP, one can construct a policy for the original POMDP which is optimal. The following observations are crucial in obtaining the equivalence of these two models: (i) any history vector of the belief-MDP is a function of the history vector of the original POMDP and (ii) history vectors of the belief-MDP is a sufficient statistic for the original POMDP.

Therefore, results developed for MDPs can be applied to the belief-MDP and so, to the POMDP. However, one should keep in mind that the correspondence between policies of the belief-MDP and the POMDP is quite complicated as one has to compute the so-called \emph{non-linear filtering equation} \cite{Her89} at each time step. Moreover, stationary policies in the belief-MDP can be history dependent in the original POMDP. Therefore, establishing structural properties of optimal policies for the belief-MDP in general does not give much information about the optimal policies for the original POMDP. However, we will see in the sequel that the belief-MDP formulation of the POMDP is useful in the finite-model approximation problem.

\section{Continuity Properties of Belief-MDPs}\label{sec2}

In this section, we first discuss the continuity properties that are satisfied by or prohibitive for the transition probability $\eta$ of the belief-MDP. Then, we derive the conditions satisfied by the components of the belief-MDP.

\subsection{On the Convergence of Probability Measures}\label{sectConvProb}
Let $\sE $ be a Borel space and let $\mathcal{P}(\sE )$ denote the family of
all probability measure on $(\sE ,\mathcal{B}(\sE ))$. A sequence $\{\mu_n\}$ is said to  converge
to $\mu\in \mathcal{P}(\sE )$ \emph{weakly} (resp., \emph{setwise}) if
\[
 \int_{\sE} g(e) \mu_n(de)  \to \int_{\sE} g(e) \mu(de)
\]
for all continuous and bounded real function $g$ (resp., for all measurable and bounded real function $g$).

For any $\mu,\nu \in \mathcal{P}(\sE )$, the \emph{total variation} norm is given by
\begin{eqnarray}
\|\mu-\nu\|_{TV}&:= & 2 \sup_{B \in {\cal B}(\sE )}
|\mu(B)-\nu(B)| \nonumber \\
 &=&  \sup_{f: \, \|f\|_{\infty} \leq 1} \bigg| \int_{\sE} f(e)\mu(de) -
\int_{\sE} f(e)\nu(de) \bigg|, \nonumber
\end{eqnarray}
where the supremum is over all measurable real $f$ such that $\|f\|_{\infty} = \sup_{e \in \sE } |f(e)|\le 1$.
A sequence  $\{\mu_n\}$ is said to  converge to $\mu\in \mathcal{P}(\sE )$ in total variation if
$\| \mu_n - \mu   \|_{TV}  \to 0$. As it is clear from the definitions, total variation convergence implies setwise convergence, which in turn implies weak convergence.


The total variation metric leads to a stringent notion for convergence. For example a
sequence of discrete probability measures on a finite-dimensional Euclidean space never converges in total
variation to a probability measure which admits a density function with respect to the Lebesgue measure. Setwise convergence also induces a topology which is not easy to work with since the space under this convergence is not metrizable \cite[p. 59]{Ghosh}. However, the space of probability measures on a Borel space endowed with the topology of weak convergence is itself a Borel space \cite{Bil99}. The bounded-Lipschitz metric $\rho_{BL}$ \cite[p.109]{Vil09}, for example, can be used to metrize this space:
\begin{align}
\rho_{BL}(\mu,\nu) \coloneqq \sup_{\|f\|_{BL}\leq1} \biggl| \int_{\sE} f(e) \mu(de) - \int_{\sE} f(e) \nu(de) \biggr|, \label{BLD}
\end{align}
where
\begin{align}
\|f\|_{BL} \coloneqq \|f\|_{\infty} + \sup_{e \neq e'} \frac{f(e) - f(e')}{d_{\sE}(e,e')}, \nonumber
\end{align}
and $d_{\sE}$ is the metric on $\sE$. Finally, the Wasserstein metric of order 1, $W_1$, can also be used for compact $\sE$ (see \cite[Theorem 6.9]{Vil09}):
\begin{align}
  W_1(\mu,\nu) =\inf_{\eta \in \mathcal{H}(\mu,\nu)} \int_{\sE \times \sE } d_{\sE}(e,e') \eta(de,de'), \nonumber
\end{align}
where $\mathcal{H}(\mu,\nu)$ denotes the set of probability measures on $\sE \times\sE $ with first marginal $\mu$ and second marginal $\nu$. Indeed, $W_1$ can also be used as an upper bound to $\rho_{BL}$ for non-compact $\sE$ since $W_1$ can equivalently be written as \cite[Remark 6.5]{Vil09}:
\begin{align}
W_1(\mu,\nu) \coloneqq \sup_{\|f\|_{Lip}\leq1} \biggl| \int_{\sE} f(e) \mu(de) - \int_{\sE} f(e) \nu(de) \biggr|, \nonumber
\end{align}
where
\begin{align}
\|f\|_{Lip} \coloneqq \sup_{e \neq e'} \frac{f(e) - f(e')}{d_{\sE}(e,e')}. \nonumber
\end{align}
Comparing this with (\ref{BLD}), it follows that
\begin{align}
\rho_{BL} \leq W_1. \label{bound}
\end{align}
This observation will be utilized for the quantization algorithms on the set of probability measures later in the paper.


\subsection{Continuity Properties under a Belief Space Formulation of POMDPs}\label{sec1sub1}

As indicated in Section~\ref{sec1}, any POMDP can be reduced to a (completely observable) MDP \cite{Yus76}, \cite{Rhe74}, whose states are the posterior state distributions or {\it beliefs} of the observer; that is, the state at time $t$ is
\begin{align}
z_t \coloneqq \sPr\{x_{t} \in \,\cdot\, | y_0,\ldots,y_t, a_0, \ldots, a_{t-1}\} \in \P(\sX). \nonumber
\end{align}
In this section, we construct the components of this belief-MDP (in particular transition probability $\eta$) under some assumptions on the components of the POMDP. Later, we establish the conditions satisfied by the components of the belief-MDP, under which we can apply approximation results in our earlier work \cite{SaYuLi16,SaYuLi17} to the belief-MDP, and so, to the original POMDP.

To this end, let $v:\sX \rightarrow [0,\infty)$ be a continuous moment function in the sense that there exists an increasing sequence of compact subsets $\{K_n\}_{n\geq1}$ of $\sX$ such that
\begin{align}
\lim_{n\rightarrow\infty} \inf_{x \in \sX \setminus K_n} v(x) = \infty. \nonumber
\end{align}
The following assumptions will be imposed on the components of the POMDP.

\begin{assumption}
\label{partial:as1}
\item [(a)] The one-stage cost function $c$ is continuous and bounded.
\item [(b)] The stochastic kernel $p(\,\cdot\,|x,a)$ is weakly continuous in $(x,a) \in \sX \times \sA$, i.e., if $(x_k,a_k) \rightarrow (x,a)$, then $p(\,\cdot\,|x_k,a_k) \rightarrow p(\,\cdot\,|x,a)$ weakly.
\item [(c)] The observation channel $r(\,\cdot\,|x)$ is continuous in total variation, i.e., if $x_k \rightarrow x$, then $r(\,\cdot\,|x_k) \rightarrow r(\,\cdot\,|x)$ in total variation.
\item [(d)] $\sA$ is compact.
\item [(e)] There exists a constant $\lambda\geq0$ such that
\begin{align}
\sup_{a \in \sA} \int_{\sX} v(y) p(dy|x,a) \leq \lambda v(x). \nonumber
\end{align}
\item [(f)] The initial probability measure $\mu$ satisfies
\begin{align}
\int_{\sX} v(x) \mu(dx) < \infty. \nonumber
\end{align}
\end{assumption}

We let
\begin{align}
\P_v(\sX) \coloneqq \biggl\{\mu \in \P(\sX): \int_{\sX} v(x) \mu(dx) < \infty \biggr\}. \nonumber
\end{align}
Note that since the probability law of $x_t$ is in $\P_v(\sX)$, by Assumption~\ref{partial:as1}-(e),(f), under any policy we have $\sPr\{x_t \in \,\cdot\, | y_0,\ldots,y_t, a_0, \ldots, a_{t-1}\} \in \P_v(\sX)$ almost everywhere. Therefore, the belief-MDP has state space $\sZ = \P_v(\sX)$ instead of $\P(\sX)$, where $\sZ$ is equipped with the Borel $\sigma$-algebra generated by the topology of weak convergence. The transition probability $\eta$ of the belief-MDP can be constructed as follows (see also \cite{Her89}). Let $z$ denote the generic state variable for the belief-MDP. First consider the transition probability on $\sX \times \sY$ given $\sZ \times \sA$
\begin{align}
R(x \in A, y \in B|z,a) \coloneqq \int_{\sX} \kappa(A,B|x',a) z(dx'), \nonumber
\end{align}
where $\kappa(dx,dy|x',a) \coloneqq r(dy|x) p(dx|x',a)$. Let us disintegrate $R$ as
\begin{align}
R(dx,dy|z,a) = H(dy|z,a) F(dx|z,a,y). \nonumber
\end{align}
Then, we define the mapping $F: \sZ \times \sA \times \sY \rightarrow \sZ$ as
\begin{align}
F(z,a,y) = F(\,\cdot\,|z,a,y) \label{eq:non_filtering}.
\end{align}
In the literature, (\ref{eq:non_filtering}) is called the `nonlinear filtering equation' \cite{Her89}.
Note that, for each $t\geq0$, we indeed have
\begin{align}
F(z,a,y)(\,\cdot\,) &= \Pr\{x_{t+1} \in \,\cdot\, | z_t = z, a_t = a, y_{t+1} = y\} \nonumber \\
\intertext{and}
H(\,\cdot\, | z,a) &= \Pr\{Y_{t+1} \in \,\cdot\, | z_t = z, a_t = a\}. \nonumber
\end{align}
Then, $\eta$ can be written as
\begin{align}
\eta(\,\cdot\,|z,a) = \int_{\sY} \delta_{F(z,a,y)}(\,\cdot\,) \text{ } H(dy|z,a), \nonumber
\end{align}
where $\delta_z$ denotes the Dirac-delta measure at point $z$; that is, $\delta_z(D) = 1$ if $z \in D$ and otherwise it is zero. Recall that the initial point for the belief-MDP is $\mu$; that is, $z_0 \sim \delta_{\mu}$, and the one-stage cost function $\tilde{c}$ of the belief-MDP is given by
\begin{align}
\tilde{c}(z,a) \coloneqq \int_{\sX} c(x,a) z(dx). \label{weak:eq8}
\end{align}
Hence, the belief-MDP is a fully-observed Markov decision process with the components
\begin{align}
\bigl(\sZ,\sA,\eta,\tilde{c}\bigr). \nonumber
\end{align}

For the belief-MDP define the history spaces $\tilde{\sH}_{0} = \sZ$ and $\tilde{\sH}_{t}=(\sZ\times\sA)^{t}\times\sZ$, $t=1,2,\ldots$ and let $\tilde{\Pi}$ denote the set of all policies for the belief-MDP, where the policies are defined in a usual manner. Let $\tilde{J}(\tilde{\pi},\xi)$ denote the discounted cost function of policy $\tilde{\pi} \in \tilde{\Pi}$ for initial distribution $\xi$ of the belief-MDP.

Notice that any history vector $\tilde{h}_t = (z_0,\ldots,z_t,a_0,\ldots,a_{t-1})$ of the belief-MDP is a function of the history vector $h_t = (y_0,\ldots,y_t,a_0,\ldots,a_{t-1})$ of the POMDP. Let us write this relation as
$i(h_t) = \tilde{h}_t$. 
Hence, for a policy $\tilde{\pi} = \{\tilde{\pi}_t\} \in \tilde{\Pi}$, we can define a policy $\pi^{\tilde{\pi}} = \{\pi_t^{\tilde{\pi}}\} \in \Pi$ as
\begin{align}
\pi_t^{\tilde{\pi}}(\,\cdot\,|h_t) \coloneqq \tilde{\pi}_t(\,\cdot\,|i(h_t)). \label{equivalence}
\end{align}
Let us write this as a mapping from $\tilde{\Pi}$ to $\Pi$: $\tilde{\Pi} \ni \tilde{\pi} \mapsto i(\tilde{\pi}) = \pi^{\tilde{\pi}} \in \Pi$. It is straightforward to show that the cost functions $\tilde{J}(\tilde{\pi},\xi)$ and $J(\pi^{\tilde{\pi}},\mu)$ are the same, where $\xi=\delta_{\mu}$. One can also prove that (see \cite{Yus76}, \cite{Rhe74})
\begin{align}
\inf_{\tilde{\pi} \in \tilde{\Pi}} \tilde{J}(\tilde{\pi},\xi) &= \inf_{\pi \in \Pi} J(\pi,\mu) \label{weak:eq7}
\end{align}
and furthermore, that if $\tilde{\pi}$ is an optimal policy for the belief-MDP, then $\pi^{\tilde{\pi}}$ is optimal for the POMDP as well. Hence, the POMDP and the corresponding belief-MDP are equivalent in the sense of cost minimization. Therefore, approximation results developed for MDPs in \cite{SaYuLi16,SaYuLi17} can be applied to the belief-MDP and so, to the POMDP.

\subsection{Strong and Weak Continuity Properties of the Belief MDP}

The stochastic kernel $\eta$ is said to be \emph{weakly} continuous if $\eta(\,\cdot\,|z_k,a_k) \rightarrow \eta(\,\cdot\,|z,a)$ weakly, whenever $(z_k,a_k) \rightarrow (z,a)$. The kernel is said to be \emph{strongly}
continuous if, for any $z \in \sZ$, $\eta(\,\cdot\,|z,a_k) \rightarrow \eta(\,\cdot\,|z,a)$ setwise, whevener $a_k \rightarrow a$. For the fully observed reduction of a partially observed MDP (POMDP), requiring strong continuity of the transition probability is in general too strong condition. This is illustrated through the following simple example.

\begin{example}\label{weak:exm1}
Consider the system dynamics
\begin{align}
x_{t+1} &= x_t + a_t, \nonumber\\
y_t &= x_t + v_t, \nonumber
\end{align}
where $x_t \in \sX$, $y_t \in \sY$, $a_t \in \sA$, and where $\sX$, $\sY$, and $\sA$ are the state, observation and action spaces, respectively, all of which we take to be $\R_{+}$ (the nonnegative real line) and the `noise' process $\{v_t\}$ is a sequence of i.i.d. random variables uniformly distributed on $[0,1]$. It is easy to see that the transition probability $p(\,\cdot\,|x,a) \coloneqq \Pr\{x_{t+1} \in \,\cdot\,| x_t =x, a_t=a\}$ is weakly continuous with respect to state-action variables $(x,a)$ and the observation channel $r(\,\cdot\,|x) \coloneqq \Pr\{y_t \in \,\cdot\,| x_t =x\}$ is continuous in total variation with respect to state variable $x$ for this POMDP. Hence, by \cite[Theorem 3.7]{FeKaZg14} the transition probability $\eta$ of the belief-MDP is weakly continuous in the state-action variables. However, the same conclusion cannot be drawn for the setwise continuity of $\eta$ with respect to the action variable as shown below.

Recall that $\eta$  is given by
\begin{align}
\eta(\,\cdot\,|z,a) = \int_{\sY} 1_{\{F(z,a,y) \in \,\cdot\,\}} H(dy|z,a), \nonumber
\end{align}
where $F(z,a,y)(\,\cdot\,) = \Pr\{x_{t+1} \in \,\cdot\, | z_t = z, a_t = a, y_{t+1} = y\}$, $H(\,\cdot\, | z,a) = \Pr\{y_{t+1} \in \,\cdot\, | z_t = z, a_t = a\}$, and $z_t \in \sZ = \P(\sX)$ is the posterior distribution of the state $x_t$ given the past observations, i.e.,
\begin{align}
z_t(\,\cdot\,) = \Pr \{x_t \in \,\cdot\,| y_0,\ldots,y_t,a_0,\ldots,a_{t-1}\}.\nonumber
\end{align}
Let us set $z = \delta_0$ (point mass at $0$), $\{a_k\} = \{\frac{1}{k}\}$, and $a = 0$. Then $a_k \rightarrow a$, but as we next show, $\eta(\,\cdot\,|z,a_k)$ does not converge to $\eta(\,\cdot\,|z,a)$ setwise.

Observe that for all $k$ and $y \in \sY$, we have $F(z,a_k,y) = \delta_{\frac{1}{k}}$ and $F(z,a,y) = \delta_0$. Define the open set $O$ with respect to the weak topology in $\sZ$ as
\begin{align}
O \coloneqq \biggl\{z \in \sZ: \biggl|\int_{\sX} g(x) \delta_1(dx) - \int_{\sX} g(x) z(dx)\biggr| < 1\biggr\}, \nonumber
\end{align}
where $g(x) = |1-x|$ if $x \in [-1,1]$ and $g(x) = 0$ otherwise. Observe that we have $F(z,a_k,y) \in O$ for all $k$ and $y$, but $F(z,a,y) \not\in O$ for all $y$. Hence,
\begin{align}
\eta(O|z,a_k) &\coloneqq \int_{\sY} 1_{\{F(z,a_k,y) \in O\}}  H(dy|z,a_k) = 1, \nonumber \\
\intertext{but}
\eta(O|z,a) &\coloneqq \int_{\sY} 1_{\{F(z,a,y) \in O\}}  H(dy|z,a) = 0, \nonumber
\end{align}
implying that $\eta(\,\cdot\,|z,a_k)$ does not converge to $\eta(\,\cdot\,|z,a)$ setwise. Hence, $\eta$ does not satisfy the strong continuity assumption.
\end{example}

The following theorem is a consequence of \cite[Theorem 3.7, Example 4.1]{FeKaZg14} and the preceding example.

\begin{theorem}\label{weak:thm6}
\item [(i)] Under Assumption~\ref{partial:as1}-(b),(c), the stochastic kernel $\eta$ for belief-MDP is weakly continuous in $(z,a)$.
\item [(ii)] If we relax the continuity in total variation of the observation channel to setwise or weak continuity, then $\eta$ may not be weakly continuous even if the transition probability $p$ of POMDP is continuous in total variation.
\item [(iii)] Finally, $\eta$ may not be setwise continuous in action variable $a$ even if the observation channel is continuous in total variation.
\end{theorem}

Part~(i) of Theorem~\ref{weak:thm6} implies that the transition probability $\eta$ of the belief-MDP is weakly continuous under Assumption~\ref{partial:as1}. However, note that continuity of the observation channel in total variation in Assumption~\ref{partial:as1} cannot be relaxed to weak or setwise continuity. On the other hand, the continuity of the observation channel in total variation is not enough for the setwise continuity of $\eta$.

The above suggest that our earlier results in \cite{SaYuLi16} and \cite{SaYuLi17}, which only require weak continuity conditions on the transition kernel of a given MDP, are particularly suitable in developing approximation methods for POMDPs (through their MDP reduction), in both quantizing the action spaces as well as state spaces.

\begin{remark}
We refer the reader to \cite[Theorem 3.2(c)]{FeKaZg14} for more general conditions implying weak continuity of the transition probability $\eta$. We also note that, in the uncontrolled setting, \cite{bhatt2000markov} and \cite{budhiraja2002invariant} have established similar weak continuity conditions (i.e., the weak-Feller property) of the non-linear filter process (i.e., the belief process) in continuous time and discrete time, respectively.
\end{remark}

\begin{example}
In this example we consider the following partially observed model
\begin{align}
x_{t+1}&=F(x_{t},a_{t},v_{t}), \nonumber \\
y_{t}&=H(x_{t},w_{t}), \text{ } t=0,1,2,\ldots \label{model}
\end{align}
where $\sX= \R^n$, $\sA \subset \R^m$, and $\sY \subset \R^d$ for some $n,m,d\geq1$. The noise processes $\{v_{t}\}$ and $\{w_t\}$ are sequences of independent and identically distributed (i.i.d.) random vectors taking values in $\sV = \R^p$ and $\sW = \R^l$, respectively, for some $p,l\geq1$, and they are also independent of each other. In this system, the continuity of $F$ in $(x,a)$ is sufficient to imply the weak continuity of the transition probability $p$, and no assumptions are needed on the noise process (not even the existence of a density is required). On the other hand, the continuity of the observation channel $r$ in total variation holds, if for any $x \in \sX$, the probability measure $r(\,\cdot\,|x)$ has a density $g(y,x)$, which is continuous in $x$, with respect to some reference probability measure $m$ on $\sY$. This follows from Scheff\'{e}'s theorem (see, e.g., \cite[Theorem 16.2]{Bil95}). For instance, this density condition holds for the following type of models:
\begin{itemize}
\item[(i)] In the first model, we have $\sY = \sW = \R^d$, $H(x,w) = H(x) + w$, $H$ is continuous, and $w$ has a continuous density $g_w$ with respect to Lebesgue measure.
\item[(ii)] In the second case, $\sY$ is countable and $r(y|x)$ is continuous in $x$ for all $y \in \sY$. Therefore, the transition probability $\eta$ of the belief space MDP, corresponding to the model in (\ref{model}), is weakly continuous.
\end{itemize}
\end{example}

Next, we derive conditions satisfied by the components of the belief-MDP under Assumption~\ref{partial:as1}. Note first that $\sZ = \bigcup_{m\geq1} F_m$, where
\begin{align}
F_m \coloneqq \biggl\{\mu \in \P_v(\sX): \int_{\sX} v(x) \mu(dx) \leq m \biggr\}. \nonumber
\end{align}
Since $v$ is a moment function, each $F_m$ is tight \cite[Proposition E.8]{HeLa96}. Moreover, each $F_m$ is also closed since $v$ is continuous. Therefore, each $F_m$ is compact with respect to the weak topology. This implies that $\sZ$ is a $\sigma$-compact Borel space. Note that by \cite[Proposition 7.30]{BeSh78}, the one-stage cost function $\tilde{c}$ of the belief-MDP, which is defined in (\ref{weak:eq8}), is in $C_b(\sZ\times\sA)$ under Assumption~\ref{partial:as1}-(a). Therefore, the belief-MDP satisfies the following conditions under Assumption~\ref{partial:as1}, which we formally state as a separate assumption.

\begin{assumption}
\label{partial:Implied}
\item [(i)] The one-stage cost function $\tilde{c}$ is bounded and continuous.
\item [(ii)] The stochastic kernel $\eta$ is weakly continuous.
\item [(iii)] $\sA$ is compact and $\sZ$ is $\sigma$-compact.
\end{assumption}

\section{Finite Model Approximations}\label{sec3}

\subsection{Finite-Action Approximation}

In this section, we consider finite-action approximation of the belief-MDP and so, the POMDP. For these equivalent models, we obtain an approximate finite-action model as follows. Let $d_{\sA}$ denote the metric on $\sA$. Since $\sA$ is assumed compact and thus totally bounded, there exists a sequence of finite sets $\Lambda_n = \{a_{n,1},\ldots,a_{n,k_n}\} \subset \sA$ such that for each $n$,
$$
\min_{i\in\{1,\ldots,k_n\}} d_{\sA}(a,a_{n,i}) < 1/n \text{ for all } a \in \sA.
$$
In other words, $\Lambda_n$ is a $1/n$-net in $\sA$. The sequence $\{\Lambda_n\}_{n\geq1}$ is used by the finite-action model to approximate the belief-MDP and the POMDP.

In \cite{SaYuLi16}, for MDPs with Borel state and action spaces, we studied the problem of approximating an uncountable action set with a finite one and had established the asymptotic optimality of finite action models for such fully observed MDPs that satisfy a number of technical conditions, in particular, the weak continuity condition of the transition kernel in state and action variables. Given the belief-MDP reduction, \cite[Theorem 3.2]{SaYuLi16} implies the following result.

\begin{theorem}
\label{weak:thm5}
Suppose Assumption~\ref{partial:as1} (and thus Assumption \ref{partial:Implied}) holds for the POMDP. Then we have
\begin{align}
\lim_{n\rightarrow\infty} |\tilde{J}^*_n(z) - \tilde{J}^*(z)| = 0 \text{  } \text{ for all $z\in\sZ$}, \nonumber
\end{align}
where $\tilde{J}^*_n$ is the discounted cost value function of the belief-MDP$_n$ with the components $\bigl( \sZ, \Lambda_n,\eta,\tilde{c} \bigr)$ and
$\tilde{J}^*$ is the discounted cost value function of the belief-MDP with components $\bigl( \sZ, \sA,\eta,\tilde{c} \bigr)$.
\end{theorem}

The significance of Theorem~\ref{weak:thm5} is reinforced by the following observation. If we let $\Pi(\Lambda_n)$ to be the set of deterministic policies of the POMDP taking values in $\Lambda_n$, then the theorem implies that for any given $\varepsilon>0$ there exists $n\geq1$ and $\pi^{*} \in \Pi(\Lambda_n)$ such that
\begin{align}
J(\pi^{*}_{\varepsilon},\mu) < \min_{\pi \in \Pi} J(\pi,\mu) + \varepsilon, \nonumber
\end{align}
where $\pi^{*}_{\varepsilon} = \pi^{\tilde{f}^{*}_n}$ (see (\ref{equivalence})) and $\tilde{f}^*_n$ is the optimal deterministic stationary policy for the belief-MDP$_n$.

\subsection{Finite-State Approximation}
\label{sec2sub0}


The finite-state model for the belief-MDP is obtained as in \cite{SaYuLi17}, by quantizing the set of probability measures $\sZ = \P_v(\sX)$; that is, for each $m$, we quantize compact set $F_m$ similar to the quantization of $\sA$ and represent the rest of the points $\sZ \setminus F_m$ by some pseudo-state. 

If $\sZ$ is compact, in the following, the index $m$ can be fixed to $m=1$.

We let $d_{\sZ}$ denote a metric on $\sZ$ which metrizes the weak topology. For each $m\geq1$, since $F_m$ is compact and thus totally bounded, there exists a sequence $\bigl(\{z_{n,i}^{(m)}\}_{i=1}^{k_n^{(m)}}\bigr)_{n\geq1}$ of finite grids in $F_m$ such that for all $n\geq1$,
\begin{align}
\min_{i\in\{1,\ldots,k_n^{(m)}\}} d_{\sZ}(z,z_{n,i}^{(m)}) < 1/n \text{ for all } z \in F_m. \label{vvv}
\end{align}
Let $\{\S_{n,i}^{(m)}\}_{i=1}^{k_n^{(m)}}$ be a partition of $F_m$ such that $z_{n,i}^{(m)} \in \S_{n,i}^{(m)}$ and
\begin{align}
\max_{z \in \S_{n,i}^{(m)}} d_{\sZ}(z,z_{n,i}^{(m)}) < 1/n \label{ccc}
\end{align}
for all $i=1,\ldots,k_n^{(m)}$. Choose any $z_{n,k_n^{(m)}+1}^{(m)} \in \sZ \setminus F_m$ which is a so-called pseudo-state and set $\S_{n,k_n^{(m)}+1} = \sZ \setminus F_m$. Let $\sZ_n^{(m)} \coloneqq \{z_{n,1}^{(m)},\ldots,z_{n,k_n}^{(m)},z_{n,k_n^{(m)}+1}^{(m)}\}$ and define function $Q_n^{(m)}:\sZ\rightarrow\sZ_n^{(m)}$ by
\begin{align}
Q_n^{(m)}(z) = z_{n,i}^{(m)} \text{ } \text{when } z \in \S_{n,i}^{(m)}. \nonumber
\end{align}
Here $Q_n^{(m)}(z)$ maps $z$ to the representative element of the partition it belongs to.

\begin{remark}
\begin{itemize}
\item [ ]
\item [(a)] Note that given $\{z_{n,i}^{(m)}\}_{i=1}^{k_n^{(m)}} \subset F_m$ that satisfies (\ref{vvv}), one way to obtain the corresponding partition $\{\S_{n,i}^{(m)}\}_{i=1}^{k_n^{(m)}}$ of $F_m$ satisfying (\ref{ccc}) as follows. Let us define function $Q_{\text{near}}: F_m \rightarrow \{z_{n,1}^{(m)},\ldots,z_{n,k_n^{(m)}}^{(m)}\}$ as
\begin{align}
Q_{\text{near}}(z) = \arg\min_{z_{n,i}^{(m)}} d_{\sZ}(z,z_{n,i}^{(m)}), \nonumber
\end{align}
where ties are broken so that $Q_{\text{near}}$ is measurable. In the literature, $Q_{\text{near}}$ is often called a nearest neighbor quantizer with respect to 'distortion measure` $d_{\sZ}$ \cite{GrNe98}. Then, $Q_{\text{near}}$ induces a partition $\{\S_{n,i}^{(m)}\}_{i=1}^{k_n^{(m)}}$ of the space $F_m$ given by
\begin{align}
\S_{n,i}^{(m)} = \{z \in F_m: Q_n^{(m)}(z)=z_{n,i}^{(m)}\}, \nonumber
\end{align}
and which satisfies (\ref{ccc}). Although one can construct, in theory, the partition using nearest neighbor sense, it is computationally difficult to find these regions when the original state space $\sX$ is uncountable.

\item [(b)] The index $n$ indicates the resolution of the quantizer that is applied to discretize the compact set $F_m$ and index $m$ emphasizes the size of the compact set $F_m$ for which quantization is applied.
\end{itemize}
%
\end{remark}


Let $\{\nu_n^{(m)}\}$ be a sequence of probability measures on $\sZ$ satisfying
\begin{align}
\nu_n^{(m)}(\S_{n,i}^{(m)}) > 0 \text{  for all  } i,n,m.  \label{compact:numeas}
\end{align}
One possible choice for $\nu_n^{(m)}$ is
\begin{align}
\nu_n^{(m)}(\,\cdot\,) = \sum_{i=1}^{k_n^{(m)}+1} \delta_{z_{n,i}^{(m)}}(\,\cdot\,). \nonumber
\end{align}
We let $\nu_{n,i}^{(m)}$ be the restriction of $\nu_n^{(m)}$ to $\S_{n,i}^{(m)}$ defined by
\begin{align}
\nu_{n,i}^{(m)}(\,\cdot\,) \coloneqq \frac{\nu_n^{(m)}(\,\cdot\,)}{\nu_n^{(m)}(\S_{n,i}^{(m)})}. \nonumber
\end{align}
The measures $\nu_{n,i}^{(m)}$ will be used to define a sequence of finite-state belief MDPs, denoted as MDP$_{n}^{(m)}$, which approximate the belief-MDP. To this end, for each $n$ and $m$ define the one-stage cost function $c_n^{(m)}: \sZ_n^{(m)}\times\sA \rightarrow [0,\infty)$ and the transition probability $p_n^{(m)}$ on $\sZ_n^{(m)}$ given $\sZ_n^{(m)}\times\sA$ by
\begin{align}\label{FinitePOMDPM}
c_n^{(m)}(z_{n,i}^{(m)},a) &\coloneqq \int_{\S_{n,i}^{(m)}} c(z,a) \nu_{n,i}^{(m)}(dz), \nonumber \\
p_n^{(m)}(\,\cdot\,|z_{n,i}^{(m)},a) &\coloneqq \int_{\S_{n,i}^{(m)}} Q_n^{(m)} \ast p(\,\cdot\,|z,a) \nu_{n,i}^{(m)}(dz),
\end{align}
where $Q_n^{(m)}\ast p(\,\cdot\,|z,a) \in \P(\sZ_n^{(m)})$ is the pushforward of the measure $p(\,\cdot\,|z,a)$ with respect to $Q_n^{(m)}$; that is,
\begin{align}
Q_n^{(m)}\ast p(y|z,a) = p\bigl(\{z \in \sZ: Q_n^{(m)}(z) = y \}|z,a\bigr), \nonumber
\end{align}
for all $y \in \sZ_n^{(m)}$. For each $n$ and $m$, we define MDP$_n^{(m)}$ as a Markov decision process with the following components: $\sZ_n^{(m)}$ is the state space, $\sA$ is the action space, $p_n^{(m)}$ is the transition probability, and $c_n^{(m)}$ is the one-stage cost function. 

Given the belief-MDP, \cite[Theorem 3.2]{SaYuLi17}\footnote{Although Theorem~3.2 in \cite{SaYuLi17} is proved under the assumption that the state space is locally compact, a careful examination of the proof reveals that $\sigma$-compactness of the state space is also sufficient to establish the result.} implies the following.

\begin{theorem}
\label{weak:thm5'}
Suppose Assumption~\ref{partial:as1} (and thus Assumption \ref{partial:Implied}) holds for the POMDP. Then we have
\begin{align}
\lim_{n,m\rightarrow\infty} |\tilde{J}(f_n^{(m)},\mu) - \tilde{J}^*(\mu)| = 0, \nonumber
\end{align}
where $f_n^{(m)}$ is obtained by extending the optimal policy of the MDP$_n^{(m)}$ to $\sZ$. Hence, by the equivalence of POMDPs and belief-MDPs, we also have
\begin{align}
\lim_{n,m\rightarrow\infty} |J(\pi^{f_n^{(m)}},\mu) - J^*(\mu)| = 0. \nonumber
\end{align}
\end{theorem}

Theorem~\ref{weak:thm5'} implies that to find a near optimal policy for the POMDP, it is sufficient to compute an optimal policy for the finite-state belief-MDP with sufficiently many states, extend this policy to the original state space of the belief-MDP, and then construct the corresponding policy for the POMDP.

In the following, we discuss explicit methods to quantize the set of probability measures on $\sX$, that is, the belief-space $\sZ$.

\section{Quantization of the Belief-Space}

An explicit construction for an application requires a properly defined metric on $\sZ$. As stated in Section \ref{sectConvProb}, one can metrize the set of probability measures defined on a Borel space under the weak topology using various distance measures. Building on this fact, in the following we present explicit methods for the quantization of $\sZ$ for the cases where $\sX$ is finite, a compact subset of a finite dimensional Euclidean space, or the finite-dimensional Euclidean space itself, and $p(\,\cdot\,|a)$ is independent of the state variable $x$.

\subsection{Construction with Finite $\sX$}\label{finite}

If the state space is finite with $|\sX|=m$, then $\sZ = \P_v(\sX) = \P(\sX)$, and $\sZ$ is a simplex in $\R^m$. In this case, Euclidean distance can be used to metrize $\sZ$.
Indeed, one can make use of the algorithm in \cite{Rez11} (see also \cite{bocherer2016optimal}) to quantize $\sZ$ in a nearest neighbor manner. To this end, for each $n\geq1$, define
\begin{align}
\sZ_n \coloneqq \biggl\{(p_1,\ldots,p_m) \in \mathbb{Q}^m: p_i = \frac{k_i}{n}, \sum_{i=1}^m k_i = n \biggr\}, \label{lattice}
\end{align}
where $\mathbb{Q}$ is the set of rational numbers and $n,k_1,\ldots,k_m \in \mathbb{Z}_{+}$. The set $\sZ_n$ is called \emph{type lattice} by analogy with the concept of \emph{types} in information theory \cite[Chp. 12]{Cover}. Then, the algorithm that computes the nearest neighbor levels can be described as follows:\\

\noindent \textbf{Algorithm.} Given $z \in \sZ$, find nearest $y \in \sZ_n:$
\begin{itemize}
\item [(1)] Compute values ($i=1,\ldots,m$)
\begin{align}
k_i' = \bigg\lfloor nz_i + \frac{1}{2} \bigg\rfloor \text{and } n' = \sum_{i=1}^m k_i'. \nonumber
\end{align}
\item [(2)] If $n' = n$ the nearest $y$ is given by $(\frac{k_1'}{n},\ldots,\frac{k_m'}{n})$. Otherwise, compute the errors
\begin{align}
\delta_i &= k_i' - n z_i, \nonumber
\intertext{and sort them}
\frac{-1}{2} \leq \delta_{i_1} \leq &\delta_{i_2} \leq \ldots \leq \delta_{i_m} \leq \frac{1}{2}. \nonumber
\end{align}
\item [(3)] Let $\Delta = n' - n$. If $\Delta > 0$, set
\begin{align}
k_{i_j} =
\begin{cases}
k_{i_j}' & \text{if } j= 1,\ldots,m-\Delta-1 \\
k_{i_j}'-1 & \text{if } j= m-\Delta, \ldots,m.
\end{cases} \nonumber
\end{align}
If $\Delta < 0$, set
\begin{align}
k_{i_j} =
\begin{cases}
k_{i_j}'+1 & \text{if } j= 1,\ldots,|\Delta| \\
k_{i_j}' & \text{if } j= |\Delta|+1, \ldots,m.
\end{cases} \nonumber
\end{align}
Then, the nearest $y$ is given by $(\frac{k_1}{n},\ldots,\frac{k_m}{n})$.
\end{itemize}

\noindent One can also compute the maximum radius of the quantization regions for this algorithm. To this end, let $d_{\infty}$ and $d_p$ denote respectively the metrics induced by $L_{\infty}$ and $L_p$ ($p\geq1$) norms on $\R^m$, which metrizes the weak topology on $\sZ$. Then, we have \cite[Proposition 2]{Rez11}
\begin{align}
b_{\infty} &\coloneqq \max_{z \in \sZ} \min_{y \in \sZ_n} d_{\infty}(z,y) = \frac{1}{n} \biggl(1 - \frac{1}{m} \biggr), \nonumber \\
b_{2} &\coloneqq \max_{z \in \sZ} \min_{y \in \sZ_n} d_{2}(z,y) = \frac{1}{n} \sqrt{\frac{a(m-a)}{m}}, \nonumber \\
b_{1} &\coloneqq \max_{z \in \sZ} \min_{y \in \sZ_n} d_{1}(z,y) = \frac{1}{n} \frac{2a (m-a)}{m}, \nonumber
\end{align}
where $a = \lfloor m/2 \rfloor$. Hence, for each $n\geq1$, the set $\sZ_n$ is an $b_j$-net in $\sZ$ with respect to $d_j$ metric, where $j \in \{\infty,2,1\}$.

\subsection{Construction with Compact $\sX$}\label{compact}

The analysis in the previous subsection shows that a finitely supported measure can be approximated through {\it type lattice}s. Thus, if compactly supported probability measures can be approximated with those having finite support, the analysis in Section~\ref{finite} yields approximately optimal policies. In the following, we assume that $\sX$ is a compact subset of $\R^d$ for some $d\geq1$. Then $\sZ \coloneqq \P_v(\sX) = \P(\sX)$ is also compact (under the weak convergence topology) and can be metrized using the  Wasserstein metric $W_1$ (here, in defining $W_1$, we use the metric on $\sX$ induced by the Euclidean norm $\|\,\cdot\,\|$) as discussed in Section \ref{sectConvProb}.


For each $n\geq1$, let $Q_n$ be some lattice quantizer \cite{GrNe98} on $\sX$ such that $\|x - Q_n(x)\| < 1/n$ for all $x \in \sX$. Set $\sX_n = Q_n(\sX)$, i.e., the output levels of $Q_n$ (note that $\sX_n$ is finite since $\sX$ is compact). Then, one can approximate any probability measure in $\sZ$ with probability measures in
\[{\cal P}(\sX_n) := \bigg\{\mu \in {\cal P}(\sX): \mu(\sX_n) = 1\bigg\}.\]
Indeed, for any $\mu \in \sZ$, we have \cite[Theorem 2.6]{kreitmeier2011optimal}
\begin{align}
\inf_{\mu' \in {\cal P}(\sX_n)} W_1(\mu,\mu') &\leq \inf_{Q: \sX \to \sX_n} \int_{\sX}  \|x-Q(x)\| \mu(dx) \nonumber \\
&\leq \int_{\sX}  \|x-Q_n(x)\| \mu(dx) \leq {1 \over n}.\nonumber
\end{align}
Once this is obtained, we can further approximate the probability measure induced by $Q_n$ via the algorithm introduced in Section~\ref{finite} with asymptotic performance guarantees. Thus, through a sequence of type lattices $\sZ_{m_n}$ as given in (\ref{lattice}) with a successively refined support set so that $\sX_n \subset \sX_{n+1}$ for $n \in \mathbb{N}$ with $m_n = |\sX_n|$, one can quantize $\sZ$ to obtain a sequence of finite state-action MDPs through (\ref{FinitePOMDPM}) leading to Theorem \ref{weak:thm5'}.

For some related properties of approximations of probability measures with those with finite support, and the relation to optimal quantization, we refer the reader to \cite{kreitmeier2011optimal}.

\subsection{Construction with Non-compact $\sX$}

Here we assume that $\sX = \R^d$ for some $d\geq1$ and that Assumption~\ref{partial:as1} holds for $v(x)=\|x\|^2$. In this case, $\sZ \coloneqq \P_v(\sX)$ becomes the set of probability measures with finite second moment and $F_m$ is the set of probability measures with finite second moments bounded by $m$. We endow here $\sZ$ with the bounded-Lipschitz metric $\rho_{BL}$, which metrizes weak convergence (see Section \ref{sectConvProb}).

We first describe the discretization procedure for $F_m$. For each $n\geq1$, set $K \coloneqq [-n,n]^d$ and let $q_n$ denote a lattice quantizer on $K$ satisfying
\begin{align}
\sup_{x \in K} \|x - q_n(x)\| < 1/n. \nonumber
\end{align}
Let $K_n$ denote the set of output levels of $q_n$; that is, $K_n = q_n(K)$. Define
%
\begin{eqnarray}
Q_n(x) = \begin{cases}   q_n(x)  \quad \quad &
	\mbox{if} \ \ x \in K \nonumber \\
 0  \quad \quad &
	\mbox{if} \ \ x \in K^c,
\end{cases} \nonumber
\end{eqnarray}

Let $\sX_n = K_n$. Then, any measure in $F_m$ can be approximated by probability measures in
\[{\cal P}(\sX_n) := \bigg\{\mu \in {\cal P}(\sX): \mu(\sX_n) = 1\bigg\}.\]
Indeed, for any $\mu \in F_m$, we have
\begin{align}
\inf_{\mu' \in {\cal P}(\sX_n)} \rho_{BL}(\mu,\mu') & \leq \inf_{\mu' \in {\cal P}(\sX_n)} W_1(\mu,\mu') \label{denk1}\\
&\leq \inf_{Q: \sX \to \sX_n} \int_{\sX}  \|x-Q(x)\| \mu(dx) \nonumber \\
&\leq \int_{\sX}  \|x-Q_n(x)\| \mu(dx) \nonumber \\
&= \int_{K}  \|x-Q_n(x)\| \mu(dx) + \int_{K^c}  \|x\| \mu(dx) \nonumber \\
&\leq \frac{1}{n} + \int_{\{\|x\| > n\}}  \|x\|^2 \mu(dx) \frac{1}{n} \nonumber \\
&\leq \frac{(1+m)}{n}
\end{align}
In the derivation above, (\ref{denk1}) follows from (\ref{bound}). Thus, $\mu$ in $F_m$ can be approximated by the $\mu_n \in \P(\sX_n)$, which is induced by the quantizer $Q_n$, with a bound $\rho_{BL}(\mu,\mu_n) \leq (1+m)/n$. Then, similar to Section~\ref{compact}, we can further approximate probability measure $\mu_n$ via the algorithm introduced in Section~\ref{finite} with again asymptotic performance guarantees by Theorem \ref{weak:thm5'}. Thus, analogous to compact case, using a sequence of type lattices $\sZ_{m_n}$ as given in (\ref{lattice}) with a successively refined support set $K_n \subset K_{n+1}$ for $n \in \mathbb{N}$ with $m_n = |\sX_n| = |K_n|+1$, one can quantize $\sZ$ to obtain a sequence of finite state-action MDPs through (\ref{FinitePOMDPM}).

\subsection{Construction for Special Models leading to Quantized Beliefs with Continuous Support}

So far, we have obtained quantized beliefs where each such quantized belief measure was supported on a finite set. For some applications, this may not be efficient and it may be more desirable to quantize the measurement space appropriately. For some further applications, a parametric representation of the set of allowable beliefs may be present and the construction of bins may be more immediate through quantizing the parameters in a parametric class.  What is essential in such models is that the bins designed to construct the finite belief-MDP correspond to balls which are {\it small} under the metrics that metrize the weak convergence as discussed in Section \ref{sectConvProb}.

\subsubsection{Quantized Measures Through Quantized Measurements}
For this section, we assume that transition probability $p(\,\cdot\,|a)$ is independent of the state variable $x$, $\sY \subset \R^{p}$ for some $p\geq1$, and Assumption~\ref{partial:as1} holds for some $v$.
In the view of Theorem~\ref{weak:thm5}, as a pre-processing set-up, we quantize the action space $\sA$, where the finite set $\sA_q$ represents the output levels of this quantizer. Hence, in the sequel, we assume that the action space is $\sA_q$.

Since $\kappa(dx,dy|a)\coloneqq r(dy|x) \otimes p(dx|a)$, we have
\begin{align}
R(x \in A, y \in B|z,a) &= \int_{\sX} \kappa(A,B|a) z(dx') \nonumber \\
&= \kappa(A,B|a), \nonumber
\end{align}
and so, the disintegration of $R$ becomes
\begin{align}
R(dx,dy|a) = H(dy|a) \otimes F(dx|a,y). \nonumber
\end{align}
Then, $\eta$ is given by
\begin{align}
\eta(\,\cdot\,|a) = \int_{\sY} \delta_{F(\,\cdot\,|a,y)}(\,\cdot\,) \text{ } H(dy|a). \nonumber
\end{align}
This implies that we can take the following set as the state space $\sZ$ of the fully-observed model instead of $\P_v(\sX)$:
\begin{align}
\sZ = \biggl\{F(\,\cdot\,|a,y): (a,y) \in \sA_q\times\sY  \biggr\}. \nonumber
\end{align}
We endow $\sZ$ with the bounded-Lipschitz metric $\rho_{BL}$. For each $n\geq1$, set $L \coloneqq [-n,n]^p$ and let $l_n$ denote a lattice quantizer on $L$ satisfying
\begin{align}
\sup_{y \in L} \|y - l_n(y)\| < 1/n. \nonumber
\end{align}
Let $\sY_n$ denote the set of output levels of $l_n$; that is, $\sY_n = l_n(L)$. Define
\begin{eqnarray}
q_n(y) = \begin{cases}   l_n(y)  \quad \quad &
	\mbox{if} \ \ y \in L \nonumber \\
 0  \quad \quad &
	\mbox{if} \ \ y \in L^c.
\end{cases} \nonumber
\end{eqnarray}
Then, finite set $\sZ_n \subset \sZ$, which is used to quantize $\sZ$, is given by
\begin{align}
\sZ_n = \biggl\{F(\,\cdot\,|a,y): (a,y) \in \sA_q\times\sY_n  \biggr\}, \nonumber
\end{align}
and the corresponding quantizer $Q_n: \sZ \rightarrow \sZ_n$ is defined as follows: given $z = F(\,\cdot\,|a,y)$, we define
\begin{align}
Q_n(z) = F(\,\cdot\,|a,q_n(y)). \nonumber
\end{align}

Note that to use $Q_n$ for constructing finite models, we have to obtain an upper bound on the $\rho_{BL}$-distance between $z$ and $Q_n(z)$. This can be achieved under various assumptions on the system components. One such assumption is the following: (i) $\sX = \R^{d}$ for some $d\geq1$, (ii) $\sY$ is compact, (ii) $p(dx|a) = g_p(x|a) m(dx)$ and $r(dy|x) = g_r(y|x) m(dy)$, (iv) $g_r$ is Lipschitz continuous with Lipschitz constant $K_r$, $g_r > \theta$ for some $\theta > 0$, and $\sup_{\{(y,x) \in \sY \times \sX\}}|g_r(y,x)| \eqqcolon \|g_r\|<\infty$. Since $\sY$ is compact, there exists $\epsilon(n)$ for each $n$ such that $\epsilon(n) \rightarrow 0$ as $n\rightarrow\infty$ and $\|y-q_n(y)\| \leq \epsilon(n)$ for all $y \in \sY$. Under the above assumptions, we have
\begin{align}
F(dx|a,y) = f(x|a,y) m(dx), \nonumber
\end{align}
where
\begin{align}
f(x|a,y) = \frac{g_r(y|x) g_{p}(x|a)}{\int_{\sX} g_r(y|x)g_{p}(x|a)m(dx)}. \nonumber
\end{align}
Since the bounded-Lipschitz metric $\rho_{BL}$ is upper bounded by the total variation distance, we obtain
\begin{align}
\rho_{BL}(z,Q_n(z)) &\leq \|z-Q_n(z)\|_{TV} \nonumber \\
&= \int_{\sX} \bigl| f(x|a,y) - f(x|a,q_n(y)) \bigr| m(dx) \nonumber \\
&\leq \frac{2\|g_r\|K_r}{\theta^2} \|y-q_n(y)\| \nonumber \\
&\leq \frac{2\|g_r\|K_r}{\theta^2} \epsilon(n).\nonumber
\end{align}
Hence, $Q_n$ is a legitimate quantizer for constructing the finite models. Section~\ref{exm_com} exhibits another example where we have such an upper bound.

\subsubsection{Construction from a Parametrically Represented Class}

For some applications, the set of belief measures can be first approximated by some parametric class of measures, where parameters belong to some low-dimensional space \cite{BrMaWiDu06,zhou2010solving,BrWi07}. For instance, in \cite{zhou2010solving}, densities of belief measures are projected onto exponential family of densities using the \emph{Kullback-Leibler (KL) divergence}, where it was assumed that projected beliefs are close enough to true beliefs in terms of cost functions. In \cite{BrWi07}, densities are parameterized by unimodal Gaussian distributions and parameterized MDP are solved through Monte Carlo simulation based method. In \cite{BrMaWiDu06}, densities are represented by sufficient statistics, and in particular represented by Gaussian distributions, and the parameterized MDP is solved through fitted value iteration algorithm. However, among these works, only the \cite{zhou2010solving} develop rigorous error bounds for their algorithms using the KL divergence and the other works do not specify distance measures to quantify parametric representation approximations.

In these methods, if the parameterized beliefs are good enough to represent true beliefs as it was shown in \cite{zhou2010solving}, then the method presented in the earlier sections (of first quantizing the state space, and then quantizing the beliefs on the state space) may not be necessary and one can, by quantizing the parameters for the class of beliefs considered, directly construct the finite belief-MDP. As noted earlier, what is essential in such methods is that the bins designed to construct the finite belief-MDP correspond to balls which are {\it small} under the metrics that metrize the weak convergence as discussed in Section \ref{sectConvProb}. This possible if the projected beliefs are proved to be close to the true beliefs with respect to some metric that generates the weak topology or with respect to some (pseudo) distance which is stronger than weak topology. For instance, since convergence in KL-divergence is stronger than weak convergence, the projected beliefs constructed in \cite{zhou2010solving} indeed satisfies this requirement. Hence, one can apply our results to conclude the convergence of the reduced model to the original model in \cite{zhou2010solving}. As noted earlier, relative entropy is a very strong pseudo-distance measure which is even stronger than total variation (by Pinsker's inequality \cite{GrayInfo}) and for being able to quantize a set of probability measures with finitely many balls as defined by such a distance measure requires very strict assumptions on the allowable beliefs and it in particular requires, typically equicontinuity conditions (see e.g. \cite[Lemma 4.3]{YukselOptimizationofChannels}). In turn, it is in general necessary to assume that transition probability and observation channel have very strong regularity conditions.

\section{Numerical Examples}\label{example}

In this section, we consider two examples in order to illustrate our results numerically. Since computing true costs of the policies obtained from the finite models is intractable, we only compute the value functions of the finite models and illustrate their converge as $n\rightarrow\infty$. We note that all results in this paper apply  with straightforward modifications for the case of maximizing reward instead of minimizing cost.

\subsection{Example with Finite $\sX$}

We consider a machine repair problem in order to illustrate our results numerically for finite state POMDPs.
In this model, we have $\sX = \sA = \sY = \{0,1\}$ with the following interpretation:
\begin{align}
x_t &= \begin{cases}
1 & \text{machine is working at time $t$} \\
0 & \text{machine is not working at time $t$,}
\end{cases} \nonumber \\
a_t &= \begin{cases}
1 & \text{machine is being repaired at time $t$} \\
0 & \text{machine is not being repaired at time $t$,}
\end{cases} \nonumber
\intertext{and}
y_t &= \begin{cases}
1 & \text{machine is measured to be working at time $t$} \\
0 & \text{machine is measured to be not working  at time $t$.}
\end{cases} \nonumber
\end{align}
There are two sources of uncertainty in the model. The first one is the measurement uncertainty. The probability that the measured state is not the true state is given by $\varepsilon$; that is,
\begin{align}
\sPr\{ y_t=0 | x_t=1\} = \sPr\{ y_t=1 | x_t=0\} = \varepsilon. \nonumber
\end{align}
In other words, there is a binary symmetric channel with crossover probability $\varepsilon$ between the state process and the observation process.

The second uncertainty comes from the repair process. In this case, $\kappa$ is the probability that the machine repair was successful given an initial `not working' state:
\begin{align}
\sPr\{ x_{t+1}=1 | x_t=0, a_t=1\} = \kappa. \nonumber
\end{align}
Finally, the probability that the machine does not break down in one time step is denoted by $\alpha$:
\begin{align}
\sPr\{ x_{t+1}=0 | x_t=1, a_t=0\} = \alpha. \nonumber
\end{align}
The one-stage cost function for this model is given by:
\begin{align}
c(x,a) = \begin{cases}
R+E & x=0 \text{ and } a=1 \\
E   & x=0 \text{ and } a=0 \\
0   & x=1 \text{ and } a=0 \\
R   & x=1 \text{ and } a=1,
\end{cases} \nonumber
\end{align}
where $R$ is defined to be the cost of repair and $E$ is the cost incurred by a broken machine. The cost function to be minimized is the discounted cost function with a discount factor $\beta$.

In order to find the approximately optimal policies, we first construct the belief space formulation of the above model. Note that the state space of the belief space model is the interval $[0,1]$. Hence, we can use uniform quantization on $[0,1]$ to obtain the finite model.

For the numerical results, we use the following parameters: $\varepsilon=0.17$, $\kappa=0.9$, $\alpha=0.9545$, and $\beta = 0.3$. We selected 20 different values for the number $n$ of grid points to discretize $[0,1]$: $n=10,20,30,\ldots,200$. The grid points are chosen uniformly. For each $n$, the finite state models are constructed as in \cite[Section 2]{SaYuLi17}.

\begin{figure}[h]
\hspace{-10pt}
\includegraphics[width=3.9in, height=2.3in]{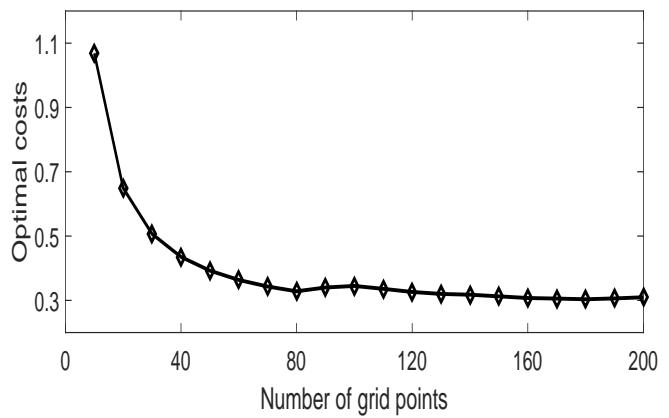}
\caption{Optimal costs of the finite models when the initial state is $x=1$}
\label{gr}
\end{figure}

Figure~\ref{gr} shows the graph of the value functions of the finite models corresponding to the different values of $n$ (number of grid points), when the initial state is $x=1$. It can be seen that the value functions converge (to the value function of the original model by \cite[Theorem 2.4]{SaYuLi17}).

\subsection{Example with Compact $\sX$}\label{exm_com}

In this example we consider the following model:
\begin{align}
x_{t+1} &= \exp\{-\theta_1 a_t + v_t\}, \text{ } t=0,1,2,\ldots \label{aux9} \\
y_t &= x_t + \xi_t, \text{ } t=0,1,2,\ldots
\end{align}
where $\theta_1 \in \R_{+}$, $x_t$ is the state at $t$, and $a_t$ is the action at $t$. The one-stage `reward' function is $u(x_t-a_t)$, where $u$ is some utility function. In this model, the goal is to maximize the discounted reward. This model is the modified and partially observed version of the population growth model in \cite[Section 1.3]{HeLa96}.

The state and action spaces are $\sX = \sA = [0,L]$, for some $L \in \R_{+}$, and the observation space is $\sY = [0,K]$ for some $K \in \R_{+}$. Since $\theta_1$ is merely a constant, by taking $[0,\frac{L}{\theta_1}]$ as our new action space, instead of dynamics in equation (\ref{aux9}) we can write the dynamics of the state as
\begin{align}
x_{t+1} = \exp\{-a_t + v_t\}, \text{ } t=0,1,2\ldots. \nonumber
\end{align}
The noise processes $\{v_{t}\}$ and $\{\xi_{t}\}$ are sequences of independent and identically distributed (i.i.d.) random variables which have common densities $g_v$ supported on $[0,\lambda]$ and $g_{\xi}$ supported on $[0,\tau]$, respectively. Therefore, the transition probability $p(\,\cdot\,|x,a)$ is given by
\begin{align}
p\bigl(D|x,a\bigr) &= \Pr \biggl\{x_{t+1} \in D \biggl| x_t=x, a_t=a\biggr\} \nonumber \\
&= \Pr \biggl\{\exp\{-a + v\} \in D\biggr\} \nonumber \\
&= \int_{D} g_v\bigl(\log(v)+a\bigr) \frac{1}{v} m(dv), \nonumber
\end{align}
for all $D\in \B(\R)$ and the observation kernel $r(\,\cdot\,|x)$ is given by
\begin{align}
r\bigl(B|x,a\bigr) &= \Pr \biggl\{y_{t} \in B \biggl| x_t=x\biggr\} \nonumber \\
&= \Pr \biggl\{ x+\xi \in B\biggr\} \nonumber \\
&= \int_{B} g_{\xi}(\xi-x)  m(d\xi), \nonumber
\end{align}
for all $B\in \B(\R)$. To make the model consistent, we must have $\exp\{-a + v\} \in [0,L]$ for all $(a,v) \in [0,L]\times[0,\lambda]$. We assume that $g_v$ and $g_{\xi}$ are uniform probability density functions; that is, $g_v = \frac{1}{\lambda}$ on $[0,\lambda]$ and $g_{\xi} = \frac{1}{\tau}$ on $[0,\tau]$. Hence, Assumption~\ref{partial:as1} holds for this model with $v(x)=1$.

In the view of Theorem~\ref{weak:thm5}, as a pre-processing set-up, we quantize the action space $\sA$, where the finite set $\sA_q = \{a_1,a_2,\ldots,a_q\}$ represents the output levels of this quantizer with $0<a_1<a_2<\ldots<a_q$. In the remainder of this example we assume that the action space is $\sA_q$.

We now obtain the stochastic kernels $H(\,\cdot\,|z,a)$ and $F(\,\cdot\,|z,a,y)$ that describe the transition probability $\eta$ of the reduced MDP. Indeed, we have
\begin{align}
H(dy|z,a) &= h(y|a) m(dy), \nonumber
\end{align}
where $h(y|a)$ is given by
\begin{align}
h(y|a) &= \int_{\sX} g_{\xi}(y-x) g_v(\log(x)+a) \frac{1}{x} m(dx) \nonumber \\
&= \int_{\sX} \frac{1}{\tau \lambda} 1_{\bigl\{\bigl[y-\tau,y\bigr] \bigcap \bigl[\exp\{-a\},\exp\{\lambda-a\}\bigr]\bigr\}}(x) \frac{1}{x} m(dx). \nonumber
\end{align}
Similarly, we have
\begin{align}
F(dx|z,a,y) &= f(dx|a,y) m(dx), \nonumber
\end{align}
where $f(x|z,a,y)$ is given by
\begin{align}
&f(x|a,y) = \frac{g_{\xi}(y-x) g_v(\log(x)+a) \frac{1}{x}}{\int_{\sX} g_{\xi}(y-x) g_v(\log(x)+a) \frac{1}{x} m(dx)} \nonumber \\
&= \frac{ 1_{\bigl\{\bigl[y-\tau,y\bigr] \bigcap \bigl[\exp\{-a\},\exp\{\lambda-a\}\bigr]\bigr\}}(x) \frac{1}{x}}{\int_{\sX} 1_{\bigl\{\bigl[y-\tau,y\bigr] \bigcap \bigl[\exp\{-a\},\exp\{\lambda-a\}\bigr]\bigr\}}(x) \frac{1}{x} m(dx)}. \label{density}
\end{align}
Hence, for any $(z,a)$, the transition probability $\eta(\,\cdot\,|z,a)$ has a support on the set of probability measures on $\sX$ having densities given by (\ref{density}).
This implies that we can take the following set as the state space $\sZ$ of the fully-observed model instead of $\P(\sX)$:
\begin{align}
\sZ = \biggl\{f(x|a,y) m(dx): (a,y) \in \sA\times\sY \text{ and } f \text{ as in (\ref{density})} \biggr\}. \nonumber
\end{align}
Note that for some $(a,y) \in \sA\times\sY$, probability density function $f(x|a,y)$ is not well-defined as $[y-\tau,y] \cap [\exp\{-a\},\exp\{\lambda-a\}] = \emptyset$, and so, we disregard these points. For the rest of the points in $\sA \times \sY$, a typical $f(x|a,y)$ can be in the following forms:
\begin{align}
f(x|a,y)&=\frac{1_{[\exp\{-a\},y]}(x) \frac{1}{x}}{\log(y)+\log(a)} \label{form1} \\
f(x|a,y)&=\frac{1_{[y-\tau,y]}(x) \frac{1}{x}}{\log(y)-\log(y-\tau)} \label{form2}  \\
f(x|a,y)&=\frac{1_{[y-\tau,\exp\{\lambda-a\}]}(x) \frac{1}{x}}{\log(\lambda-a)-\log(y-\tau)}. \label{form3}
\end{align}
For each $n$, let $q_{n}$ denote the uniform quantizer on $\sY$ having $n$ output levels; that is,
\begin{align}
q_{n}&:\sY \rightarrow \{y_1,\ldots,y_{n}\} \eqqcolon \sY_n \subset \sY \nonumber \\
\intertext{where $y_j = (j-\frac{1}{2})\Delta_n$, $j=1,\ldots,n$, and}
q_n^{-1}(y_j) &= \biggl[ y_j - \frac{\Delta_n}{2}, y_j + \frac{\Delta_n}{2} \biggr), \nonumber
\end{align}
where $\Delta_n = \frac{K}{n}$. We define
\begin{align}
\sZ_n \coloneqq \biggl\{f(x|a,y) m(dx) \in \sZ: (a,y) \in \sA_q\times\sY_n \biggr\}. \nonumber
\end{align}
Then, the quantizer $Q_n: \sZ \rightarrow \sZ_n$, which is used to construct the finite model, is defined as follows: given $z = f(x|a,y)m(dx)$, we define
\begin{align}
Q_n(z) = f(x|a,q_n(y)) m(dx).\nonumber
\end{align}

To be able to use $Q_n$ for constructing finite models, we need to obtain an upper bound on the $\rho_{BL}$-distance between $z$ and $Q_n(z)$. To this end, let $\theta > 0$ be a small constant such that $\exp\{-a_q\}-\theta > 0$.

Suppose that the density of $z$ is in the form of (\ref{form2}); that is, $y-\tau > \exp\{-a\}$ and $y < \exp\{\lambda-a\}$ for some $a \in \sA_q$. Let $y_n \coloneqq q_n(y)$, $\gamma \coloneqq \frac{1}{\log(y)-\log(y-\tau)}$, and $\gamma_n \coloneqq \frac{1}{\log(y_n)-\log(y_n-\tau)}$. Since $\frac{1}{\log(y)-\log(y-\tau)}$ is a continuous function of $y$ and $[\exp\{-a_q\}+\tau,\exp\{\lambda-a_1\}]$ is compact, there exists $\epsilon(\Delta_n)$ which is independent of $z$ such that $|\gamma-\gamma_n|<\epsilon(\Delta_n)$ and $\epsilon(\Delta_n) \rightarrow 0$ as $n\rightarrow\infty$. We also suppose that $y_n < y$ without loss of generality. Then, for sufficiently large $n$, by using the inequality $\log(x)\leq x-1$, we obtain
\begin{align}
\rho_{BL}&(z,Q_n(z)) \leq \|z-Q_n(z)\|_{TV} \nonumber \\
&= \int_{\sX} \bigl| f(x|a,y) - f(x|a,q_n(y)) \bigr| m(dx) \nonumber \\
&= \int_{\sX} \biggl| 1_{[y-\tau,y]}(x) \frac{1}{x} \gamma - 1_{[y_n-\tau,y_n]}(x) \frac{1}{x} \gamma_n \biggr| m(dx) \nonumber \\
&= \int_{y_n-\tau}^{y-\tau} \frac{1}{x}\gamma_n m(dx) + \int_{y-\tau}^{y_n} \biggl| \frac{1}{x} \gamma - \frac{1}{x} \gamma_n \biggr| m(dx) \nonumber \\
&\phantom{xxxxxxxxxxxxxxxxxxxxxxxxxxx}+ \int_{y_n}^{y} \frac{1}{x}\gamma m(dx) \nonumber \\
&= \gamma_n \log(\frac{y-\tau}{y_n-\tau}) + |\gamma-\gamma_n| \log(\frac{y_n}{y-\tau}) + \gamma \log(\frac{y}{y_n}) \nonumber \\
&\leq 2 K_1 \frac{\Delta_n}{K_2} + L_1 \epsilon(\Delta_n), \nonumber
\end{align}
where
\begin{align}
K_1 &\coloneqq \frac{1}{\log(\frac{\exp\{\lambda-a_1\}+\tau}{\exp\{\lambda-a_1\}})} \nonumber \\
K_2 &\coloneqq \exp\{-a_q\}-\theta \nonumber \\
L_1 &\coloneqq \log(\frac{\exp\{-a_q\}+\tau}{\exp\{-a_q\}}). \nonumber
\end{align}
Hence, $\rho_{BL}(z,Q_n(z)) \rightarrow 0$ as $n\rightarrow\infty$. Similar computations can be done for $z \in \sZ$ of the form (\ref{form1}) and (\ref{form3}). This implies that $Q_n$ is a legitimate quantizer to construct finite-state models.

For the numerical results, we use the following values of the parameters:
\begin{align}
\lambda = 1, \text{ } \tau=0.5, \text{ }\beta = 0.2. \nonumber
\end{align}
The utility function $u$ is taken to be quadratic function; i.e., $u(t) = t^2$. As a pre-processing set-up, we first uniformly discretize the action space $\sA$ by using the 20 grid points. Then, we selected 99 different values for the number $n$ of grid points to discretize the state space $\sZ$ using the quantizer $Q_n$, where $n$ varies from $29$ to $1436$.

\begin{figure}[h]
\hspace{-10pt}
\includegraphics[width=3.9in, height=2.3in]{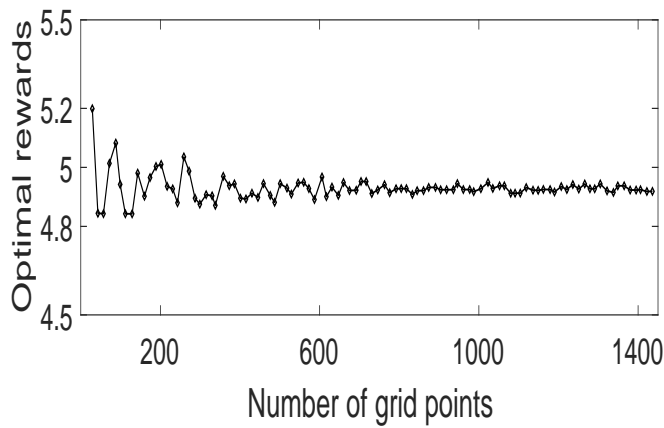}
\caption{Optimal rewards of the finite models when the initial state is $x=2$}
\label{gr2}
\end{figure}

We use the value iteration algorithm to compute the value functions of the finite models. The simulation was implemented by using MATLAB and it took 411.75 seconds using an HP EliteDesk 800 desktop computer. Figure~\ref{gr2} displays the graph of these value functions corresponding to the different values for the number of grid points, when the initial state is $x=2$. The figure illustrates that the value functions of the finite models converge (to the value function of the original model by \cite[Theorem 2.4]{SaYuLi17}).

\section{Concluding Remarks}\label{conc}
We studied the approximation of discrete-time partially observed Markov decision processes under the discounted cost criterion. An essential observation was that establishing strong continuity properties for the reduced (belief) model is quite difficult for general state and action models, whereas weak continuity can be established under fairly mild conditions on the transition kernel of the original model and the measurement equations. This allowed us to apply our prior approximation results \cite{SaYuLi16,SaYuLi17}, developed under weak continuity conditions, to partially observed models.

In particular, \cite{SaYuLi16,SaYuLi17} developed finite model approximation for both the discounted and the average cost criteria. However, for the belief-MDP, the regularity conditions (i.e., drift inequality and minorization condition) imposed on the transition probability in \cite{SaYuLi16,SaYuLi17} for the average cost problem do not hold in general. Indeed, as we observed, even setwise continuity is prohibitive for the transition probability of the belief-MDP. Therefore, we have restricted our attention to the discounted cost case. Extending the analysis to average cost problems is a future task.

\section{Acknowledgements}\label{ack}
We are grateful to Prof. Eugene Feinberg for general discussions related to the regularity properties of POMDPs. We also acknowledge Marwan Galal, Mark Gaskin, Ian Harbell and Daniel Kao (all senior students in the Mathematics and Engineering program at Queen's University) for their help with the numerical computations. We are also thankful to three anonymous reviewers who have provided extensive constructive feedback.

\bibliographystyle{IEEEtran}
\bibliography{references,SerdarBibliography}

\end{document}